\documentclass[longauth]{aa} % for the long lists of affiliations 
\usepackage{graphicx}
\usepackage[varg]{txfonts}
\usepackage{natbib}

\bibpunct{(}{)}{;}{a}{}{,}

\begin{document}
   \title{The Gaia-ESO Survey: processing of the FLAMES-UVES spectra\thanks{Based on observations made with the ESO/VLT, at 
   Paranal Observatory, under program 188.B-3002 (The Gaia-ESO Public Spectroscopic Survey)}}

   \author{G.~G.~Sacco\inst{\ref{inst1}}\and L.~Morbidelli\inst{\ref{inst1}}\and E.~Franciosini\inst{\ref{inst1}}
   \and E.~Maiorca\inst{\ref{inst1}} \and S.~Randich\inst{\ref{inst1}} \and A.~Modigliani\inst{\ref{inst2}} \and G.~Gilmore\inst{\ref{inst3}} 
    \and M.~Asplund\inst{\ref{inst4}} \and J.~Binney\inst{\ref{inst5}} \and 
   P.~Bonifacio\inst{\ref{inst6}} \and J.~Drew\inst{\ref{inst7}} \and S.~Feltzing\inst{\ref{inst8}} \and
   A.~Ferguson\inst{\ref{inst9}} \and R. Jeffries\inst{\ref{inst10}} \and G. Micela\inst{\ref{inst11}} \and
   I.~Negueruela\inst{\ref{inst12}} \and T. Prusti\inst{\ref{inst13}} \and H.-W. Rix\inst{\ref{inst14}} \and
   A.Vallenari\inst{\ref{inst15}} \and E.~Alfaro\inst{\ref{inst16}} \and C.~Allende Prieto\inst{\ref{inst17},\ref{inst17b}} \and
   C.~Babusiaux\inst{\ref{inst6}} \and T.~Bensby\inst{\ref{inst18}} \and R.~Blomme\inst{\ref{inst19}} \and
  A.~Bragaglia\inst{\ref{inst20}} \and E.~Flaccomio\inst{\ref{inst11}} \and P.~Francois\inst{\ref{inst6}} \and
   N.~Hambly\inst{\ref{inst21}} \and M.~Irwin\inst{\ref{inst3}} \and S.~Koposov\inst{\ref{inst3}} \and A.~Korn\inst{\ref{inst22}} \and 
   A. Lanzafame\inst{\ref{inst23}} \and E.~Pancino\inst{\ref{inst20},\ref{inst24}} \and A.~Recio-Blanco\inst{\ref{inst25}} \and
    R.~Smiljanic\inst{\ref{inst26}, \ref{inst28}} \and S. Van Eck\inst{\ref{inst27}} \and N. Walton\inst{\ref{inst3}} \and
    M.~Bergemann\inst{\ref{inst3}} \and M.~T.~Costado\inst{\ref{inst16}} \and P.~de Laverny\inst{\ref{inst25}} 
    \and U.~Heiter\inst{\ref{inst22}} \and V.~Hill\inst{\ref{inst25}} 
    \and A.~Hourihane\inst{\ref{inst3}} \and R.~Jackson\inst{\ref{inst10}} \and P.~Jofre\inst{\ref{inst3}} \and
    J.~Lewis\inst{\ref{inst3}} \and K.~Lind\inst{\ref{inst3}} \and C.~Lardo\inst{\ref{inst20}} \and 
    L.~Magrini\inst{\ref{inst1}} \and T.~Masseron\inst{\ref{inst3}} \and L.~Prisinzano\inst{\ref{inst11}} \and
    C.~Worley \inst{\ref{inst3}}   
     }

   \institute{INAF-Osservatorio Astrofisico di Arcetri, Largo E. Fermi, 5, 50125, Firenze, Italy\label{inst1}
   \and European Southern Observatory, Karl-Schwarzschild-Strasse 2, 85748 Garching bei München, Germany\label{inst2}
   \and Institute of Astronomy, University of Cambridge, Madingley Road, Cambridge CB3 0HA, United Kingdom\label{inst3}
   \and Research School of Astronomy \& Astrophysics, Australian National University, Cotter Road, Weston Creek, ACT 2611, Australia\label{inst4}
   \and Rudolf Peierls Centre for Theoretical Physics, Keble Road, Oxford, OX1 3NP, United Kingdom\label{inst5}
   \and GEPI, Observatoire de Paris, CNRS, Universit\'e Paris Diderot, 5 Place Jules Janssen, 92195 Meudon, France\label{inst6}
   \and Centre for Astronomy Research, University of Hertfordshire, Hatfield, Hertfordshire, AL10 9AB United Kingdom \label{inst7}
   \and Lund Observatory, Department of Astronomy and Theoretical Physics, Box 43, SE-221 00 Lund, Sweden \label{inst8}
   \and Institute of Astronomy, University of Edinburgh, Blackford Hill, Edinburgh EH9 3HJ, United Kingdom\label{inst9}
   \and Astrophysics Group, Research Institute for the Environment, Physical Sciences and Applied Mathematics, Keele University, Keele, Staffordshire ST5 5BG, United Kingdom\label{inst10}
   \and INAF - Osservatorio Astronomico di Palermo, Piazza del Parlamento 1, 90134, Palermo, Italy \label{inst11}
   \and Departamento de F\'{i}sica, Ingenier\'{i}a de Sistemas y Teor\'{i}a de la Se$\tilde{\rm n}$al, Universidad de Alicante, Apdo. 99, 03080, Alicante, Spain\label{inst12}
   \and ESA, ESTEC, Keplerlaan 1, Po Box 299 2200 AG Noordwijk, The Netherlands\label{inst13}
   \and Max-Planck Institut f\"{u}r Astronomie, K\"{o}nigstuhl 17, 69117 Heidelberg, Germany\label{inst14}
   \and INAF - Padova Observatory, Vicolo dell'Osservatorio 5, 35122 Padova, Italy\label{inst15}
   \and Instituto de Astrof\'{i}sica de Andaluc\'{i}a-CSIC, Apdo. 3004, 18080, Granada, Spain\label{inst16}
   \and Instituto Astrof\'{\i}sica de Canarias, 38205, La Laguna, Tenerife, Spain\label{inst17}
   \and Universidad de La Laguna, Depart. de Astrof\'{\i}sica, 38206, La Laguna, Tenerife, Spain\label{inst17b}
   \and Lund Observatory, Department of Astronomy and Theoretical Physics, Box 43, SE-221 00 Lund, Sweden\label{inst18}
   \and Royal Observatory of Belgium, Ringlaan 3, 1180, Brussels, Belgium\label{inst19}
   \and INAF - Osservatorio Astronomico di Bologna, via Ranzani 1, 40127, Bologna, Italy\label{inst20}
   \and Institute of Astronomy, University of Edinburgh, Blackford Hill, Edinburgh EH9 3HJ, United Kingdom\label{inst21}
   \and Department of Physics and Astronomy, Division of Astronomy and Space Physics, Uppsala University, Box 516, SE-75120 Uppsala, Sweden\label{inst22}
   \and Dipartimento di Fisica e Astronomia, Sezione Astrofisica, Universit\'{a} di Catania, via S. Sofia 78, 95123, Catania, Italy\label{inst23}
   \and ASI Science Data Center, Via del Politecnico SNC, 00133 Roma, Italy\label{inst24}
   \and Laboratoire Lagrange (UMR7293), Universit\'e de Nice Sophia Antipolis, CNRS,Observatoire de la C\^ote d'Azur, CS 342293, F-06304 Nice cedex 4, France\label{inst25}
   \and Department for Astrophysics, Nicolaus Copernicus Astronomical Center, ul. Rabia\'{n}ska 8, 87-100 Toru\'{n}, Poland\label{inst26}
   \and Institut d'Astronomie et d'Astrophysique, Universit\'{e} libre de Brussels, Boulevard du Triomphe, 1050 Brussels, Belgium\label{inst27}
   \and European Southern Observatory, Karl-Schwarzschild-Str. 2, 85748 Garching bei M\"unchen, Germany \label{inst28}
   }

   \date{Received ; accepted }

% \abstract{}{}{}{}{} 
% 5 {} token are mandatory
 
  \abstract
  % context heading (optional)leave it empty if necessary  
   {The Gaia-ESO Survey is a large public spectroscopic survey that aims to derive
    radial velocities and fundamental parameters of about 10$^5$ Milky Way stars in the field and
   in clusters. Observations are carried out with the multi-object optical spectrograph FLAMES,
   using simultaneously the medium resolution (R$\sim$20,000) GIRAFFE spectrograph and 
   the high resolution (R$\sim$47,000) UVES spectrograph.
   In this paper, we describe the methods and the software used for the data reduction,
   the derivation of the radial velocities, 
   and the quality control of the FLAMES-UVES spectra. 
   Data reduction has been performed using a workflow specifically developed for this project. 
   This workflow runs the ESO public pipeline optimizing the data reduction
   for the Gaia-ESO Survey, performs automatically sky subtraction, barycentric correction and 
   normalisation, and calculates radial velocities and a first guess of the rotational velocities.
   The quality control is performed using the output parameters from the ESO pipeline,
   by a visual inspection of the spectra and by the analysis of the signal-to-noise ratio
   of the spectra.
   Using the observations of the first 18 months, specifically targets observed multiple times at different epochs, 
   stars observed with both GIRAFFE and UVES, and observations of radial velocity standards, we estimated the precision and the 
   accuracy of the radial velocities. The statistical error on the
   radial velocities is $\sigma\sim$0.4 $\rm km~s^{-1}$ and is mainly due to uncertainties 
   in the zero point of the wavelength calibration. However, we found a systematic bias
   with respect to the GIRAFFE spectra ($\sim0.9~ \rm km~s^{-1}$) and to the radial velocities of the standard
   stars ($\sim0.5~ \rm km~s^{-1}$) retrieved from the literature. This bias will be corrected in the future data releases, when
   a common zero point for all the setups and instruments used for the survey will be established.
  } 

   \keywords{Methods:data analysis, Techniques: spectroscopic, Techniques: radial velocities, Surveys, Stars:general  }

   \maketitle
%
%________________________________________________________________

\section{Introduction}

The Gaia-ESO Survey is a large public spectroscopic survey aimed at deriving 
radial velocities (RVs), stellar parameters, and abundances 
of about 10$^5$ Milky Way stars in the field and 
in clusters \citep{Gilmore:2012, Randich:2013}. The observations started at the end
of 2011 and are expected to last for about 5 years. 

The observations are carried out with the multi-object optical spectrograph
FLAMES \citep{Pasquini:2002}. This instrument is located at the 
Nasmyth focus of the UT2 at the Very Large Telescope (VLT) and
is composed of a robotic fibre positioner equipped with two sets of 132 and 8 fibres, which feed
the optical spectrographs GIRAFFE (R$\sim$20,000) and UVES (R$\sim$47,000), respectively.
A good fraction of the spectra ($\sim$3500 from GIRAFFE and $\sim$300 from UVES) 
observed during the first six months (December 2011-June 2012) 
have been released and are available at the webpage http://www.eso.org/sci/observing/phase3/data\_releases.html.

Twenty working groups (WGs) are in charge of the workflow, which includes all steps from the 
selection of the targets to be observed to the derivation of the stellar parameters. 
This paper describes the methods and software used for the reduction of the 
FLAMES-UVES spectra and for the derivation of RVs and rotational velocities
projected along the line of sight ($v \sin i$). We will focus on
the spectra gathered during the first 18 months of observations (from December 2011 to June 2013). 
Whilst all other steps of the workflow are performed in a distributed fashion, namely
several nodes analyse the same data, and the results are finally made homogeneous by the WG coordinators,  
the work discussed in this paper has been carried out by one team based at INAF-Osservatorio Astrofisico di Arcetri.

The content of the paper is summarized as follows: in Sect. 2 we briefly describe the 
target sample and the observations; in Sect. 3 we explain 
the procedures used for the data reduction of the FLAMES-UVES spectra; in Sect. 4 we 
describe the methods used to derive RVs and $v\sin i$; in Sect. 5 we 
summarize our quality control procedure; in Sect. 6 we list the final products of our 
WG; and in Sect. 7 a summary of the paper is provided.

\section{Target sample and observations}

The target sample of the Gaia-ESO Survey includes a large variety of stars (dwarfs and giants)
with spectral types ranging from O to M and expected metallicities [Fe/H] from about -2.5 to +0.5 dex.
The setup centered at 580 nm (480-680 nm) is used for all FLAMES-UVES
observations with the exception of the early-type stars, that are observed with the setup centered 
at 520 nm (420-620 nm), and of the targets selected for calibration and testing purposes,
which are observed with all FLAMES-UVES setups, including the one centered at 860 nm (760-960 nm).

Typical exposure times of observing blocks (OBs) observed with the 580 setup are either 
1200 or 3000 s, with the exception of the targets observed for calibration, especially the very bright ones, 
that are observed for a very short time to avoid saturation. However, an OB can be repeated multiple times and a target can be 
included in different OBs, so the time spent on a single target can be longer. 
Each OB is divided into two separate exposures to help
the removal of cosmic rays. A very short exposure including an arc lamp spectrum is executed 
between the two main exposures to provide a nearly simultaneous calibration of the GIRAFFE
observations, with the exception of the observations carried out with the HR21 setup (848-900 nm),
which are calibrated using sky emission lines.

Note that the determination of very precise RVs is
not the primary goal of the UVES observations and the acquisition of a simultaneous 
arc lamp spectrum can be performed only losing one science target, so 
to maximize the number of UVES targets, we perform the wavelength calibration using
the arc lamp frame acquired during the day. 
Usually, all the 8 fibres (6 for the 520 nm setup) are allocated 
with at least one fibre on an empty position of the sky to allow 
the subtraction of the sky emission. 
The FLAMES fibre positioner is equipped with two plates\footnote{Since September 2012, one of the 
fibre on plate 2 has been dismissed due to a hardware problem. So on this plate we can allocate only 
seven targets including the sky.}, that are both used to reduce
observing overheads; in fact, fibres can be positioned on one plate, while the other one is used for observing.  

\begin{figure*}
   \centering
   \includegraphics[width=15 cm]{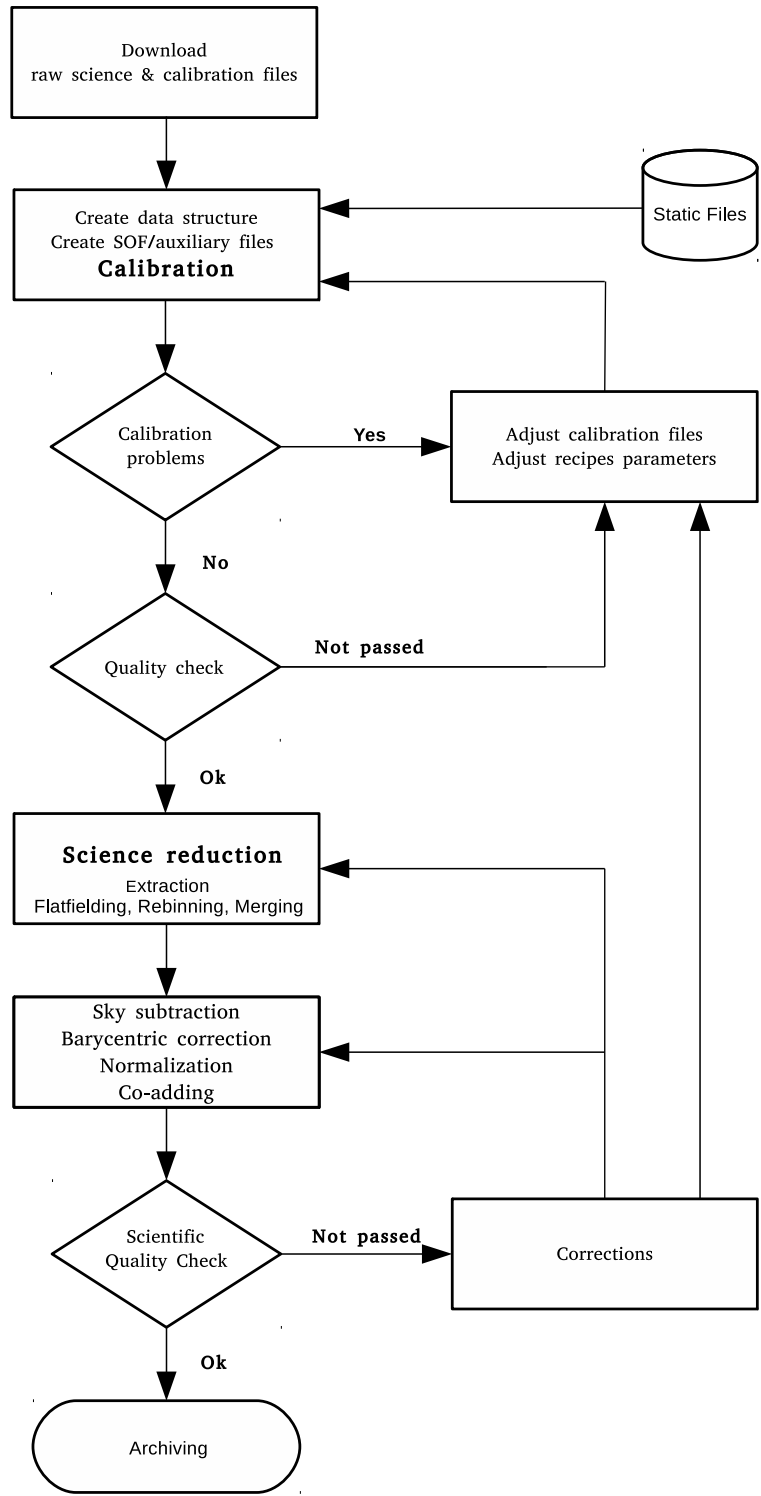}
      \caption{Flowchart describing the workflow for the reduction of the FLAMES-UVES spectra for the Gaia-ESO 
      Survey.}
         \label{fig:flowchart}
   \end{figure*}

\section{Data volume and structure \label{sect:data_red}}

The survey started on 31 December 2011. All the observations are carried out 
in Visitor mode and are divided in runs of 5-6 nights, with a monthly frequency.
During the first 18 months, we performed 17 observing runs for a total of 87 observing nights. 
We processed a total of 6971 FLAMES-UVES spectra
of 1611 stars, which have been internally released to the WGs in charge of the spectral analysis. 
All stars have been observed with the 580 setup, while only 27 stars have been observed also
with the 860 setup. Due to data reduction problems discussed in the next section, stars observed with the
520 setup have not yet been released.

We organize our data reduction flow on single night basis, namely we use the same 
set of calibration frames for all the observations carried out during the same night with the same setup
and the same plate. A set of calibrations is composed of five bias frames, nine full slit flat-fields,
three fibre flat-fields, two frames for the format definition, and one frame for the wavelength calibration.
A typical observing night includes 8-10 OBs ($\sim$24-30 exposures), which are associated to
different sets of calibrations (from one to six), depending on the setups and the plates used 
for the observations.
However, the amount of science frames taken during a night strongly depends on
the type of targets and the weather conditions. For the calibrations, we use frames 
taken in daytime soon after the observing night, with the exception of specific cases, where
our quality control identifies a poor quality of some calibration frames.
In these cases we use the calibration frames suitable for our observations, which are
closest in time.

\section{Data reduction \label{sect:data_red}}

We process the spectra with a data flow software composed of a combination 
of public software (i.e. ESO public pipeline, IRAF\footnote{IRAF is distributed by the 
National Optical Astronomy Observatories, which are operated by the Association of 
Universities for Research in Astronomy, Inc., under cooperative agreement with the National
Science Foundation}, Pyraf\footnote{Pyraf is a product of the Space Telescope Science Institute, which is operated by AURA for NASA.}) 
and a set of bash, IDL, and python scripts developed by our team. The whole data flow 
can be divided into three main parts: a) the reduction of raw frames to produce wavelength-calibrated 
spectra, which is performed by a set of scripts running the public ESO pipeline \citep{Modigliani:2004, Modigliani:2012} 
within a workflow optimized for the survey observing strategy; 
b) the basic steps of the data analysis (i.e. sky-subtraction, barycentric correction, 
normalization and co-adding) and the derivation of RV, $v\sin i$ and binarity flags, which are performed using a set of IDL and 
IRAF/Pyraf \citep{Tody:1986, Tody:1993} scripts;
c) the scientific quality control of the final products.

\begin{figure*}
   \centering
   \includegraphics[width=18 cm]{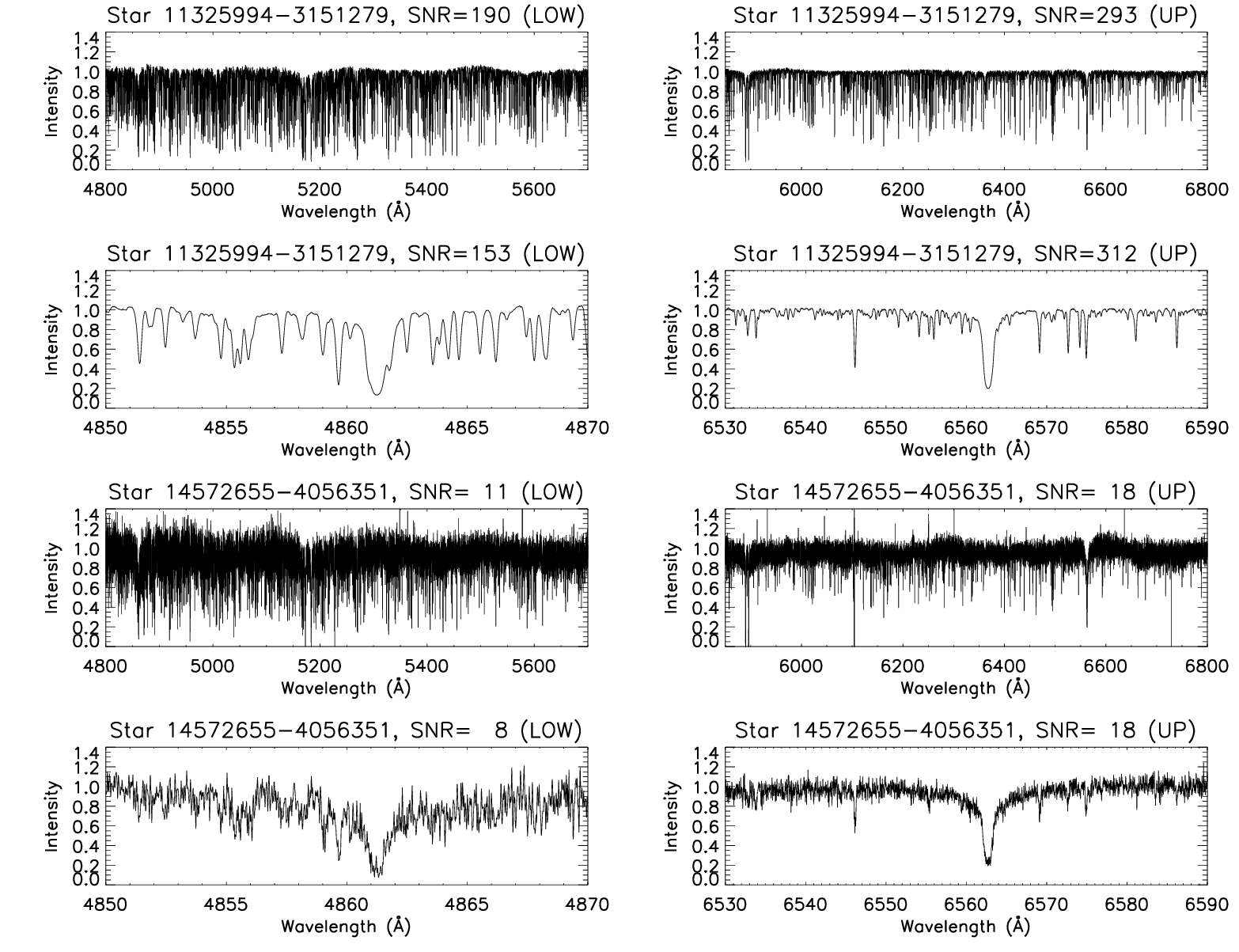}
      \caption{Examples of co-added and normalised spectra of the Gaia-ESO Survey. 
      The panels in the first two rows show different wavelength ranges of a spectrum with
      high signal-to-noise ratio (star 11325994-3151279). 
      Specifically, the first row shows the entire lower (left panel) and upper (right panel) spectra, while
      the second row shows two smaller portions of the spectrum around the H$\beta$ (left panel) and 
      H$\alpha$ (right panel) lines. The panels in the last two rows show the same ranges of
     a spectrum with low signal-to-noise ratio (star 14572655-4056351). On the top of each panel
     is reported the median of the signal-to-noise ratio per pixel for the spectral range plotted 
     in the panel.}
         \label{fig:spectra}
   \end{figure*}

The FLAMES-UVES data reduction pipeline was originally developed as a MIDAS based 
pipeline \citep{Mulas:2002} and later ported to ESO Common Pipeline Library. 
It consists of a chain of seven recipes, which perform the following steps:

\begin{figure*}
   \centering
   \includegraphics[width=18 cm]{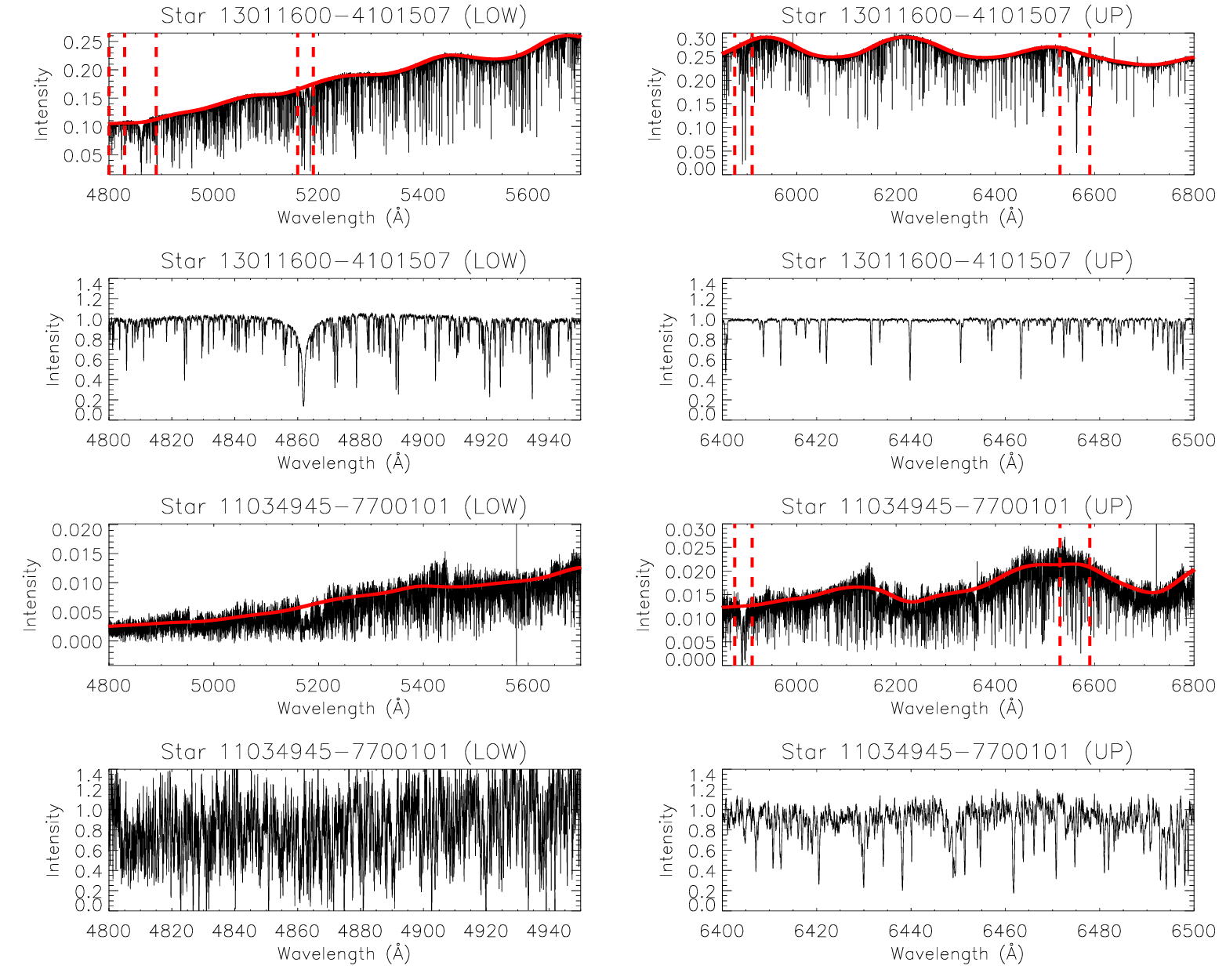}
      \caption{Examples of the results obtained with the normalisation procedure used for processing the spectra of
      the Gaia-ESO survey. The four top panels show a typical spectrum (star 13011600-4101507), while the four bottom panels
      show one of the case (star 1103495-7700101) when our procedure does not perform efficiently. The continuous 
      red lines overplotted on the spectra before the normalisation
      show the profile of the continuum calculated by the pipeline, while the vertical dashed lines indicate the wavelength intervals masked
      to avoid overnormalisation of strong lines.}
         \label{fig:normspectra}
   \end{figure*}

\begin{itemize}

\item combining raw bias frames into a master bias;

\item  computing guess tables with order 
positions on the detector, using a physical model of the instrument
and a raw frame which is acquired by 
illuminating a fibre with a line emission lamp;

\item computing a more accurate table with
order positions from a raw frame taken with the calibration
fibre illuminated by a continuum lamp;

\item creating the master slit flat-field frame by combining several
long slit exposures taken with a continuum lamp; 

\item determining the fibre order table
and constructing several frames needed to extract a science fibre
frame, using input fibre frames obtained by illuminating the fibres with
a continuum source;

\item determining the wavelength dispersion coefficients and constructing a wavelength 
calibration table from a frame where all the fibres are illuminated
by an arc line calibration lamp;

\item extracting the science frame producing 
the reduced spectra and their variances. The output spectra and variances are given
in three different formats: a) single echelle order spectra 
before the wavelength calibration, b) single echelle order 
spectra after the wavelength calibration, c) a wavelength 
calibrated spectrum created by merging all the echelle orders.

\end{itemize}

It is worth noting that the FLAMES-UVES detector is the mosaic of two CCDs, which cover the
redder and the bluer part of the spectral format. 
The ESO pipeline processes data from each CCD independently, and 
provides for each target two output files, each covering half of the full wavelength range of the setup. 
We keep the two spectral ranges separated for the whole 
data flow, so all the subsequent steps of the spectra processing described in this paper 
are performed independently for each spectral range. 
In the rest of the paper, we will refer to the two spectral ranges as lower and upper spectrum.

These recipes can be executed via the command line interface ESOREX, the graphic interface Gasgano
or the Reflex workflow. However, ESOREX requires additional software to classify the files and to 
organise the workflow, while Gasgano and Reflex are designed for interactive data reduction, and thus are
not the best choice reduction of very large datasets.
Therefore, we built a workflow to manage efficiently the ESOREX based data reduction process and
perform the quality control on the calibrations. Specifically, our workflow performs
automatically the following operations: a) classifies the raw data files in categories 
(e.g. science, bias, flat-field and so on); b) groups the calibration raw files in sets of
calibration frames, according to the setups and the plates used for the observations; 
c) executes the cascade of recipes for each set of calibration frames; 
d) associates the output of the calibration recipes to each science raw frame;
e) executes the last recipe on each science raw file; f) produces
tables and plots for quality control. 

A schematic flowchart of the workflow is shown in Fig. \ref{fig:flowchart}. 
A script starts the calibration procedure by executing consecutively the set of recipes, using the ESOREX
command line interface. At the end of the calibration phase the output files are
checked in order to detect problems and assess quality. If quality problems are detected
(a detailed description of quality control procedure and of the quality issues affecting 
the data is given in Sect. \ref{sect:qc}) 
or some recipes did not complete successfully, it is necessary to
check where the problem originates, fix it by selecting different
values of the key parameters or by choosing different calibration files,
and start again the workflow. 
Only after all calibrations
are reduced properly, we run the recipe to reduce the science frame.

Since the first period of observations, we experienced problems with 
the wavelength calibration and the definition of the order positions 
of the frames acquired with the 520 setup.
For this reason, data taken with the setup at 520 nm are not part of the first releases.
However, the ESO data reduction team recently solved this issue and released a new version
of the pipeline in November 2013.  Several tests have shown that the quality of the spectra reduced with this new pipeline are 
equivalent for all setups. Therefore, all spectra will be delivered in the next release.

After all the recipes of the ESO pipeline have been executed and 
the quality of the calibration has been assessed we perform the following operations:

\begin{enumerate}

\item We subtract the sky background spectrum from the stellar spectra. 
The sky background spectrum is usually acquired by one fibre pointing  
toward an empty position of the field of view. If more than one fibre is
used to sample the sky emission, we compute the median of the sky emission spectra
(or the average if they are only two).

\item We shift all the sky-subtracted spectra to an heliocentric 
reference frame, using the IRAF task RVCORRECT to
calculate the velocity shift due to the Earth rotation, the 
motion of the Earth center about the Earth-Moon barycenter and
the motion of the Earth-Moon barycenter about the center of
the Sun.

\item After the end of each run, 
we co-add and store in a single file all the spectra from different exposures of the same OB
and/or different OBs with the same configuration, which have been observed during the same night.
Very short exposures taken for the wavelength calibration of the GIRAFFE spectra are not co-added.
Before each data release, we produce a final spectrum and a single final file for each star. 
This final spectrum is the sum of all the spectra of the same star acquired during 
the whole survey. Note that
a specific target can be part of different OBs observed in different 
nights or different runs, so some of the final spectra are the sum of 
multi-epoch observations. As explained in more detail in the next section, 
all the candidate binaries are flagged. We do not perform the subtraction of the 
telluric absorption features before co-adding, therefore, when strong telluric features 
affect multi-epoch observations, the final co-added spectrum maybe affected by multiple telluric features.
To handle this problem and any other issues related to the variability of the spectra,
we include in the file with the final co-added spectrum also the original single-epoch spectra,
before co-adding. 

\item We normalise the merged spectra by dividing them by
a function, which describes the stellar continuum emission 
convolved with the FLAMES-UVES instrumental response. To
derive this function, we divide the spectrum in 30 bins,
compute the median in each bin, and then fit the obtained values
with a spline function, using an iterative sigma-clipping
to remove absorption and emission features. Strong lines (e.g.
Balmer lines) are masked before the calculation of the 
continuum function to avoid over-normalisation.
Some examples of spectra before and after the normalisation and
of the function used to define the continuum are shown in Fig. \ref{fig:normspectra}.  
As shown in the four bottom panels of the figure, our procedure may not work for very noisy spectra
and late-type stars. Furthermore, in many cases the procedure for the normalisation of the spectra
needs to be tuned on the basis of method used for the spectral analysis. Therefore, both spectra before and after the 
normalisation are internally released, and the teams performing the spectral analysis can re-normalize them, before deriving
stellar parameters and abundances.

\item We calculate RVs, $v\sin i$ and associated flags
to assess the quality of these measurements, as described in the next section.

\end{enumerate}

The first three steps discussed above (sky subtraction, 
heliocentric correction and co-adding) are applied to both the single
orders and the merged spectra, while only the merged spectra are
normalised. Variances of the spectra are propagated across these
steps following basic error propagation theory. 
As shown in Fig. \ref{fig:flowchart}, after these four steps have been
completed, we perform a scientific quality control (see section \ref{sect:qc}).

\begin{figure}
   \centering
   \includegraphics[width=7.5 cm]{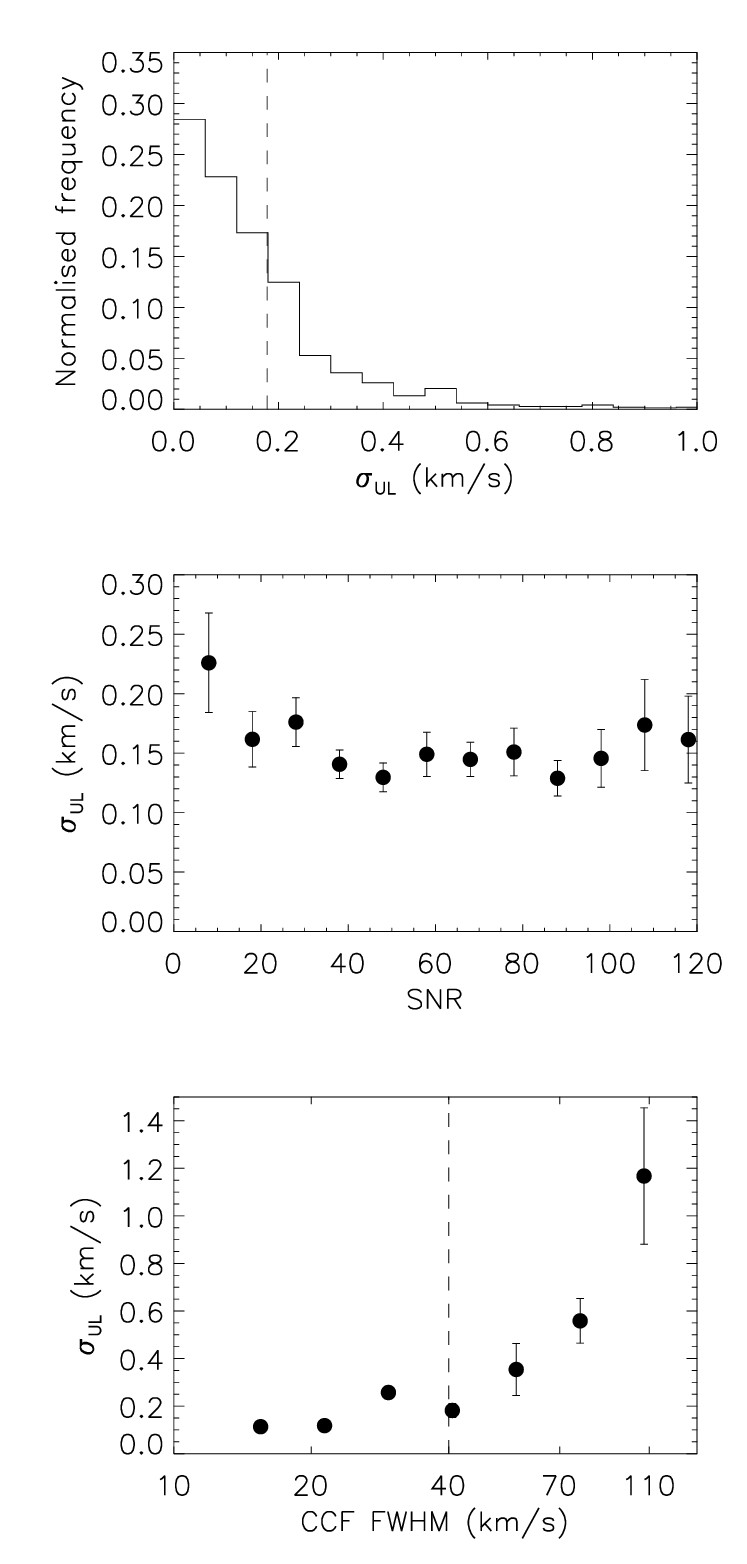}
      \caption{Empirical estimate of the errors on radial velocities. 
      {\bf Top panel:} the normalised frequency distribution of empirical uncertainties
      ($\sigma_{UL}\sim |RV_U-RV_L|/\sqrt{2}$) derived from the difference 
      between velocities measured from the upper ($RV_U$) and lower ($RV_L$) spectrum of all the stars
      observed with the 580 setup. The dashed line shows the position of the 68$^{th}$ percentile
      of the distribution. 
      {\bf Middle and bottom panels:} the same empirical uncertainties
      binned as a function of the SNR of the spectra (middle panel) 
     and of the full width half maximum of the CCF (bottom panel). Error bars on each bin
     are equal to $\sigma_{bin}/\sqrt{N_{bin}}$, where $\sigma_{bin}$ and $N_{bin}$ are 
     the standard deviation and the total number of values for each bin, respectively. The number of
     values per bin is not constant, but it ranges      
     from $\sim$200 (in the central bins, $SNR\sim 40-50$) to $\sim50$ (bins of the lowest and highest SNR) in the middle plot, 
     and from $\sim$900 ($CCF_{FWHM}\sim20~\rm km~s^{-1}$) to $\sim$20  ($CCF_{FWHM}\sim110\rm~km~s^{-1}$) in the bottom plot. 
     }
         \label{fig:rv1}
   \end{figure}

\section{Radial velocities and rotational broadening \label{sect:radial_velocities} }

We derive the stellar RVs by cross-correlating each spectrum
with a grid of synthetic template spectra. Our grid is a subsample of the library
produced by \cite{de-Laverny:2012} and is composed of 36 synthetic spectra
convolved at the FLAMES-UVES spectral resolution. It covers seven effective temperatures
($\rm T_{eff}$= 3100, 4000, 5000, 6000, 7000, 8000 K), three surface gravities (log(g) = 2.5, 4.0, 5.0) and 
two values of metallicities ([Fe/H]= 0.0, -1.0).

Each spectrum is cross-correlated with all the spectra of the grid, using the IRAF task FXCOR 
\citep{Fitzpatrick:1993}, masking the Balmer lines (H$\alpha$ and H$\beta$) and regions of the spectra with strong telluric lines. 
To derive the RV, we select the cross-correlation function (CCF) with the highest peak and fit the
peak with a Gaussian function to derive its centroid. This procedure fails for early-type stars with an effective temperature
above the highest temperature of our grid, which are characterized by the presence of no, or very few, absorption
lines other than the Balmer lines. A WG dedicated to the analysis of the early-type stars
will provide RVs for these stars, by a best-fitting procedure with an appropriate grid of templates.
Details on this procedure will be given in a forthcoming paper.

To estimate the precision of the RVs derived by our pipeline, 
we used the differences between RVs measured from the lower ($RV_L$) and upper spectra ($RV_U$), 
which are measured independently by the pipeline.
Assuming identical uncertainties on RVs from the two wavelength ranges, and 
since there is no systematic offset between lower and upper spectra ($median (RV_U-RV_L)=\rm 0.007~km~s^{-1}$), 
the statistical error on the RVs derived by our pipeline is
$\sigma_{UL}=|RV_U-RV_L|/\sqrt{2}$. The distribution of
these empirical errors for the stars observed with the 580 setup\footnote{Since no spectra
observed with the 520 setup have been released and only 27 benchmark stars have been
observed with the 860 setup, we limit our
analysis of the RV precision and accuracy to the spectra observed with the 580 setup.} during the
first 18 months of the survey is shown in the top panel of Fig. \ref{fig:rv1}. The distribution
deviates from a Gaussian due to extended wings associated to a small fraction of spectra which are affected 
by large errors (e.g., very low signal-to-noise ratio spectra, fast rotators, spectroscopic binaries).
Excluding these outliers, the statistical error on RV 
is equal to the 68$^{th}$ percentile rank of the distribution ($\sigma=0.18\rm ~km~s^{-1}$).

We use the empirical error based on the RV differences between the upper and lower spectrum to
investigate how the precision of the RVs depends on the signal-to-noise ratio (SNR) and $v \sin i$.
The middle and bottom panels of Fig. \ref{fig:rv1} show the empirical error binned by SNR and full width 
half maximum of the CCF ($CCF_{FWHM}$), respectively. The latter is correlated with
$v \sin i$ and can be directly measured. A small fraction of RVs ($\sim3\%$) have been excluded after 
a sigma-clipping applied to each bin. 
The error on RV shows almost no dependence on the SNR, while it strongly increases in fast rotators. Specifically,
the error is constant for a $CCF_{FWHM}$ smaller than $\sim$40 $\rm km~s^{-1}$ ($v \sin i\sim15~\rm km~s^{-1}$) 
and increases above. Due to the low statistics, it is not possible to determine a relation between
the error and the $CCF_{FWHM}$ for fast rotators, therefore we include in our products a RV quality
flag calculated using the maximum of the CCF and the $CCF_{FWHM}$. Specifically, we flag all the stars
with $CCF_{FWHM}>40\rm ~km~s^{-1}$ and a maximum of the CCF lower than 0.3. 
Errors on RVs may also depend on the stellar metallicity [Fe/H].
However, the number of metal-poor stars is not high enough to study the relation between 
$\sigma_{UL}$ and [Fe/H]. To obtain a first estimation on this source of error, we calculate the 68$^{th}$ percentile rank of the distribution of 
$\sigma_{UL}$ for the metal-poor stars ([Fe/H]$<-1.0$ dex, 68$^{th}$ rank $\sim0.24\rm ~km~s^{-1}$) and 
metal-rich stars ([Fe/H]$>+0.1$ dex, 68$^{th}$ rank $\sim0.14\rm ~km~s^{-1}$). 
The latter is slightly higher suggesting that measurements of RV for metal poor stars are less precise.

\begin{figure}[htb]
   \centering
   \includegraphics[width=7.5 cm]{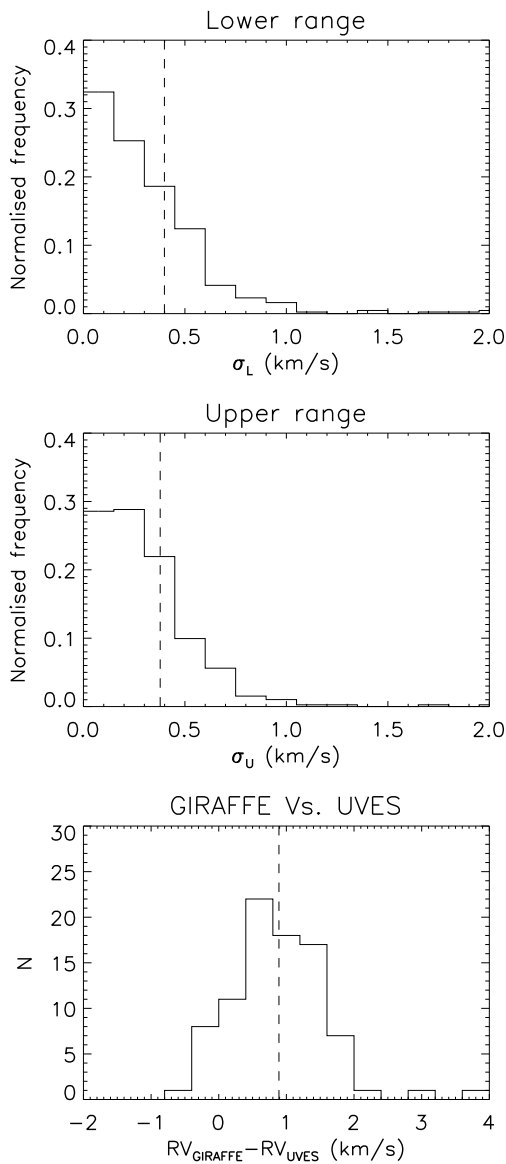}
      \caption{Errors due to the zero point of wavelength calibration. The top and the
      middle panels show the normalised frequency distributions of empirical uncertainties 
      ($\sigma\sim |\Delta RV|/\sqrt{2}$) derived from stars observed multiple times in different
      epochs for the lower and upper wavelength range, respectively. The dashed lines show the position
      of the 68\% percentile. The bottom panel shows the distribution of the differences between
      radial velocities observed with both the UVES 580 setup and the GIRAFFE HR15N setup.
      The dashed line shows the position of the median of the differences.    }
         \label{fig:rv2}
   \end{figure}

Since the upper and the lower spectrum are calibrated using the same arc lamp, our approach for 
the error estimate does not take into account the error due to the variations of 
the zero point of the wavelength calibration. In order to estimate this source of uncertainty, we used spectra of targets 
observed multiple times in different epochs. The top (lower wavelength range) and middle (upper wavelength range)
plots in Fig. \ref{fig:rv2} show the distributions of an empirical error defined as above ($\sigma=|\Delta RV|/\sqrt{2}$),
where $|\Delta RV|$ is the difference between two observations of the same target performed
in different nights. The two distributions are much wider than the distribution reported in
the top panel of Fig. \ref{fig:rv1} (the 68th percentile ranks $\sigma_U=0.38\rm ~km~s^{-1}$ and $\sigma_L=0.40\rm ~km~s^{-1}$ for the 
lower and upper ranges, respectively), which proves that the variations of the zero point of the wavelength
calibration are the main source of uncertainty. Therefore, we will adopt $\sigma \sim 0.4\rm ~km~s^{-1}$ as 
typical error for the RVs derived from the FLAMES-UVES spectra\footnote{We quoted a statistical error
$\sigma=0.6\rm~km~s^{-1}$ for the first internal data releases. This preliminary and more conservative 
estimate of the statistical error may have been used in some of the first science verification papers of the Gaia-ESO
consortium.} of the Gaia-ESO Survey.

\begin{figure}[!htb]
   \centering
   \includegraphics[width=7.5 cm]{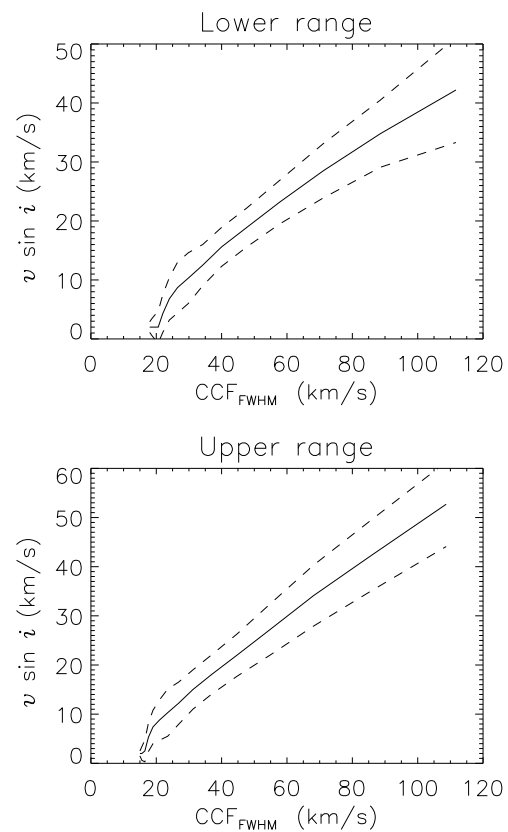}
      \caption{The continuous line describes the relations between $v \sin i$ and the $CCF_{FWHM}$
      used to derive $v \sin i$ from the lower (top panel) and upper wavelength range
      for the stars observed with the 580 setup. Dashed lines describe the same relations
      plus/minus the error bars.}
         \label{fig:rotvel}
   \end{figure}

To estimate the accuracy of our RV measurements, we observed 18 RV standards from the
catalogue developed for the calibration of the Gaia Radial Velocity Spectrograph \citep{Soubiran:2013}.
Furthermore, we compare RVs from UVES spectra with RVs from GIRAFFE spectra, for a sample of stars in
common between the two instruments. GIRAFFE observations are carried out with several setups 
HR03 (403-420 nm), HR05A (434-459 nm), HR06 (454-478 nm), HR09B (514-536 nm), HR10 (534-562 nm), 
HR14A (631-670), HR15N  (647-679 nm), and HR21 (848-900 nm). 
We consider only stars observed with the HR15N (647-679 nm) setup to avoid
the complications associated to the cross-calibration
of different GIRAFFE setups and because most of the stars in common between the 
two instruments belongs to clusters, that have been observed only with this setup.

\begin{table*}
\caption{Radial velocities of Gaia standard stars.}
\label{tab:RV_STAND}
\centering
\begin{tabular}{cccccccc}
\hline\hline
Star       &    RA         & DEC      & B-V        &RV\tablefootmark{a} & RV$_{GES}$\tablefootmark{b}              & RV$_{GES}$-RV & |RV$_{GESU}$-RV$_{GESL}$|/$\sqrt{2}$ \\
           &   (J2000)     &  (J2000) &  (mag)     &($\rm km~s^{-1}$) & ($\rm km~s^{-1}$)       & ($\rm km~s^{-1}$) & ($\rm km~ßs^{-1}$) \\
\hline
  HIP017147  &03:40:21.7  &-03:12:59.3  &   0.55  & 120.400$\pm$0.0066  &  119.99  &  -0.41  &   0.02 \\
  HIP026973  &05:43:26.4  &-47:49:24.6  &   0.86  &  26.600$\pm$0.0057  &   26.30  &  -0.30  &   0.04 \\
  HIP029295  &06:10:34.7  &-21:51:46.5  &   1.49  &   4.892$\pm$0.0088  &    3.56  &  -1.33  &   0.45 \\
  HIP031415  &06:35:03.0  &-12:36:26.2  &   0.51  &  -7.479$\pm$0.0115  &   -7.71  &  -0.23  &   0.11 \\
  HIP032045  &06:41:43.0  &-33:28:11.1  &   1.05  &  40.722$\pm$0.0065  &   39.95  &  -0.77  &   0.11 \\
  HIP032103  &06:42:23.4  &-61:13:31.1  &   0.75  &  27.167$\pm$0.0062  &   27.36  &   0.19  &   0.04 \\
  HIP038747  &07:55:58.2  &-09:47:49.7  &   0.67  &  -8.002$\pm$0.0071  &   -8.39  &  -0.39  &   0.28 \\
  HIP045283  &09:13:44.8  &-42:18:37.0  &   0.58  &  39.451$\pm$0.0053  &   38.54  &  -0.91  &   0.19 \\
    HD88725  &10:14:08.2  & 03:09:08.2  &   0.61  & -21.976$\pm$0.0052  &  -22.49  &  -0.52  &   0.07 \\
   HIP51007  &10:25:11.2  &-10:13:44.4  &   1.46  &  21.758$\pm$0.0064  &   21.01  &  -0.75  &   0.61 \\
  HIP058345  &11:57:56.9  &-27:42:19.9  &   1.13  &  48.605$\pm$0.0088  &   47.47  &  -1.14  &   0.25 \\
   HIP65859  &13:29:59.1  & 10:22:47.2  &   1.49  &  14.386$\pm$0.0087  &   13.79  &  -0.60  &   0.37 \\
  HIP077348  &15:47:24.2  &-01:03:48.6  &   0.79  &   1.907$\pm$0.0114  &    2.42  &   0.52  &   0.21 \\
  HIP104318  &21:07:56.4  & 07:25:58.7  &   0.69  &   4.910$\pm$0.0059  &    3.91  &  -1.00  &   0.12 \\
  HIP105439  &21:21:23.7  &-51:45:08.6  &   0.65  &  17.322$\pm$0.0062  &   17.33  &   0.01  &   0.25 \\
   HD204587  &21:30:02.2  &-12:30:34.0  &   1.26  & -84.533$\pm$0.0092  &  -85.58  &  -1.05  &   0.51 \\
  HIP108065  &21:53:41.7  &-28:40:12.3  &   0.73  & -41.660$\pm$0.0100  &  -42.30  &  -0.64  &   0.07 \\
  HIP113576  &23:00:16.7  &-22:31:28.2  &   1.38  &  16.138$\pm$0.0095  &   17.04  &   0.90  &   0.89 \\
\hline
\end{tabular}
\tablefoot{Radial velocities of stars included in the catalogue of standards for the
 calibration Gaia Radial Velocities Spectrometer \citep{Soubiran:2013} \\
 \tablefoottext{a}{Radial velocity from \citealt{Soubiran:2013}.}\\
 \tablefoottext{b}{Average of the radial velocities of the lower and upper spectrum derived by pipeline used to process FLAMES-UVES spectra.}
}
\end{table*}

The list of RV standards observed during the first 18 months is reported in Table \ref{tab:RV_STAND}, 
while the bottom panel of Fig. \ref{fig:rv2} shows the distribution of the differences between the 
RVs measured with the two instruments. The RVs from the UVES spectra reported both in the table and in the plot
are the average between the values calculated from the upper and the lower wavelength ranges.
There is a systematic offset with respect to both the RVs of standards ($<RV_{UVES}-RV_{ST}>=-0.47\pm0.14~\rm km~s^{-1}$) 
and the RVs measured by the GIRAFFE pipeline for the spectra taken with the HR15N setup 
($<RV_{UVES}-RV_{GIRAFFE}>=-0.85\pm0.07 ~\rm km~s^{-1}$).
The offsets have the same sign, but the latter is significantly larger.
This suggests that the RVs derived from both instruments need a more accurate zero point calibration.
We are currently investigating how to improve the wavelength calibration, using the sky emission lines or 
the telluric features included in the spectra, and we are carrying out a comparison of RVs measured from all the different setups
used for GIRAFFE and UVES to define a unique zero point for the Gaia-ESO Survey.  
Note that, as discussed by \cite{Worley:2012}, such small errors on the RVs 
do not affect the derivation of the stellar parameters.

The $CCF_{FWHM}$ is correlated with the rotational velocity, so we can use 
the $CCF$ computed for the determination of the RVs to estimate $v \sin i$. 
However, the $CCF_{FWHM}$ also depends on the stellar parameters,
which are not determined by our workflow, so the values of $v \sin i$ derived are only 
first guess estimations, that can be improved after the stellar parameters have been derived.  
To derive the relations between $v \sin i$ and the $CCF_{FWHM}$  shown in Fig. \ref{fig:rotvel}, 
we created a set of rotationally broadened synthetic spectra, by convolving 
all the template spectra used for deriving the RVs with the rotational profile derived by \cite{Gray:2008}.
For each template, we created 12 spectra with $v \sin i$ ranging between 2 and 60 $\rm km~s^{-1}$ and
distributed on a logarithmic scale. Then, we run our procedure for the determination of the RVs on the
whole set of rotationally broadened templates and from the results we derived a relation between 
$CCF_{FWHM}$ and $v \sin i$.  This relation was inverted to derive
$v \sin i$ from the $CCF_{FWHM}$ calculated for the observed stars. Errors on $v \sin i$ have been also
calculated from the dispersion of the $CCF_{FWHM}$ at fixed $v \sin i$ (see Fig. \ref{fig:rotvel}).

\begin{figure*}[htb]
   \centering
   \includegraphics[width=14 cm]{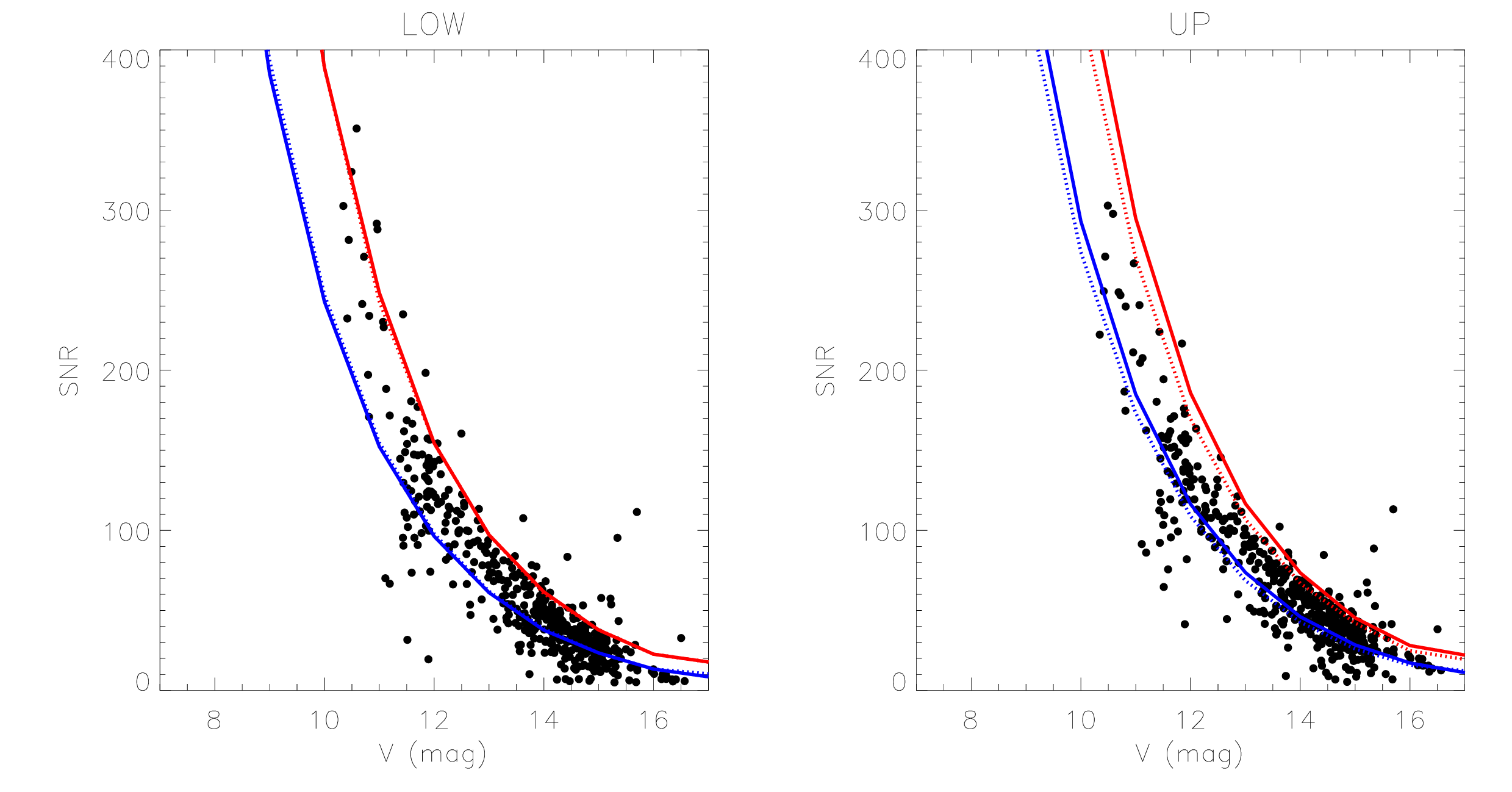}
   \includegraphics[width=14 cm]{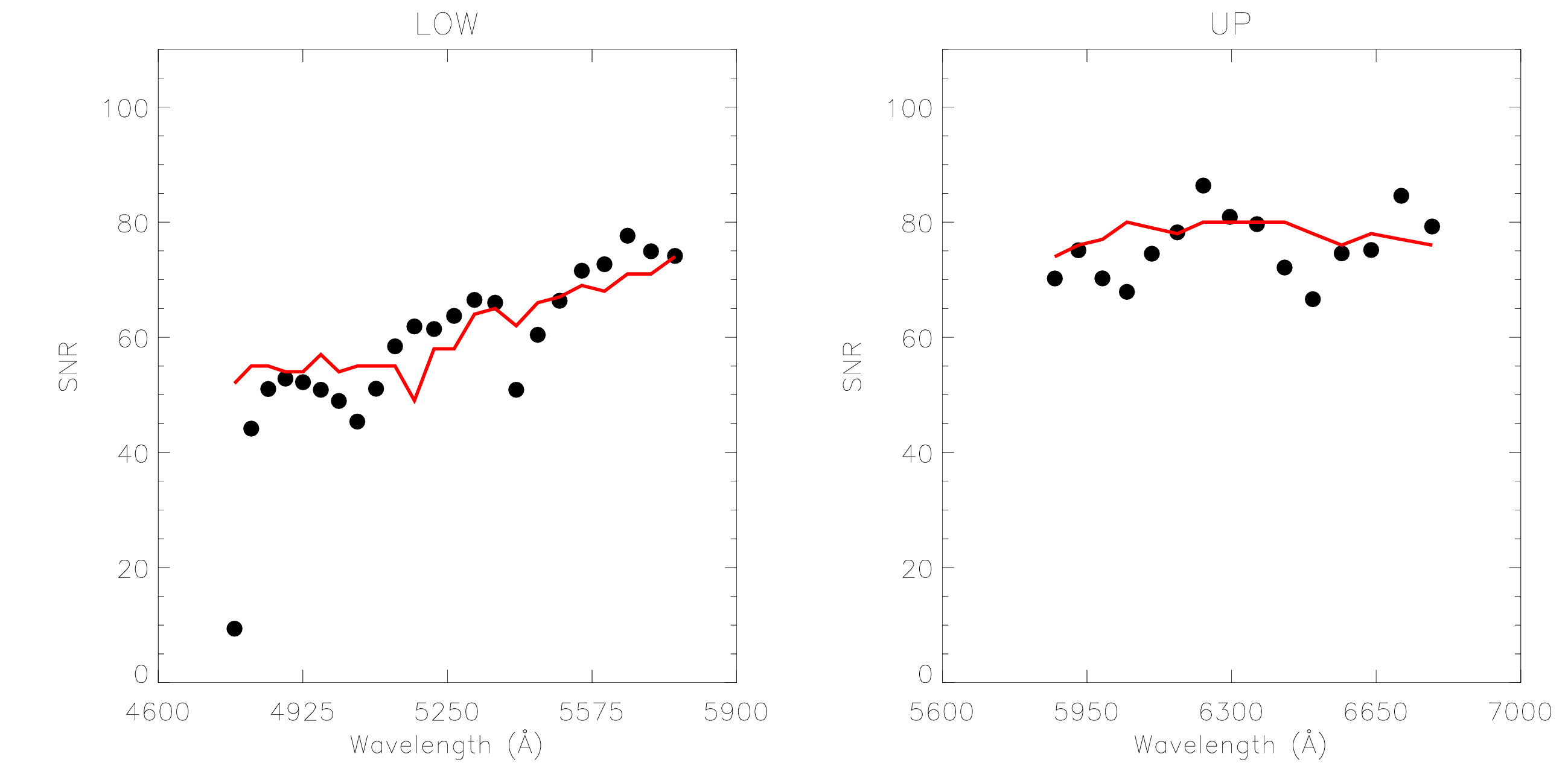}
      \caption{{\bf Top panels:} Comparison between predicted (continuous lines) and measured (black dots) SNRs 
      for the stars observed during the first 18 months of the survey, with known V magnitude from the literature or 
      from public archives. The upper and lower spectra are shown in the left and right panels,
      respectively. The lines show the predicted SNR calculated by the ESO exposure time
      calculator (vers. 5.0.1) in the 5470-5520 \AA~(left panel) and the 6250-6320 \AA~(right panel) 
      spectral ranges, considering an exposure time of 3000 s, airmass of 1.2, the input spectra of 
      a K2 (continuous line) and a G0 (dotted line) star and two values of seeing $0.66\arcsec$ (red line) and 
      $1.57\arcsec$ (blue line), which correspond to the 10$^{th}$ and 90$^{th}$ percentile ranks of the seeing measured 
      during the observations. The black dots represent
      the SNRs of the observed spectra measured in the same wavelength range using the DER SNR algorithm (see Sect.
      \ref{sect:qc}) and scaled to the exposure time of 3000 s, assuming that the SNR is photon limited 
      (i.e., SNR$\propto \sqrt{exposure~time}$).
      {\bf Bottom panels:} Comparison between predicted (continuous red line) and measured SNRs (black dots) 
      as a function of wavelength for the star 18280330+0639516. The lower and upper wavelength ranges are shown
      in the left and in the right panels, respectively. }
         \label{fig:SNR_vs_Mag}
   \end{figure*}

As said before, in order to have only one spectrum per star at the end of the survey, we co-added
all the repeated observations. However, to avoid errors in the spectral analysis due
to the presence of double-lined spectroscopic binaries or to single-lined spectroscopic
binaries observed multiple times, we include in our final products two binarity flags:
a) we perform a visual inspection of the $CCFs$ computed before co-adding multi-epoch observations, 
and flag a star as a candidate double-lined spectroscopic binary, if the $CCFs$
are characterized by the presence of more than one peak or a single peak with strong asymmetries; b)
we classify a star as a single-lined spectroscopic binary if the median absolute deviation
of multi-epoch repeated measurements of the RV is larger than twice the error on RV calculated as discussed above.

\section{Quality control \label{sect:qc}}

We perform a quality control of the calibration frames and the spectra to check
that the data reduction software is working correctly and
that problems during the fiber allocation process do not 
affect the final quality of the spectra.
Specifically, our procedure for quality control consists of three steps:

\begin{enumerate} 

\item We store several output parameters from the ESO pipeline, which 
allow us to assess the stability of the BIAS frames (e.g., BIAS level),
the accuracy of the table which defines the spectral format (e.g., root mean square of the 
shifts between the spectral format derived from a calibration frame and an analytical model of the spectral format), and 
the precision of the wavelength calibration (e.g., number of lines used and
root mean square of the residuals of the wavelength solution). 
Whenever, during the data reduction process and the following steps
of the quality control, we come across a problem (e.g. crash of the pipeline,
artifacts in the spectra), we use these parameters to investigate the origin of the
problem. In particular, we analyse if the parameters assume anomalous values 
with respect to the typical values observed during the survey or if they follow  a trend.
For this analysis, we also used the ESO Health Check Monitor for FLAMES-UVES
(available at the website http://www.eso.org/observing/dfo/quality/UVES/ qc/qc1.html), 
which allows us to compare the quality of the calibration frames used for processing the data of the 
survey with the typical quality of the FLAMES-UVES calibration frames.  

\item We perform a visual quality control of all the spectra to check
the presence of anomalies in the spectrum, like artificial noise, gaps and ripples. 
When we find such problems, we investigate if they originate
from the science frame, the calibration frames or from the pipeline, and
we evaluate possible actions to be taken, in some cases in collaboration
with the ESO data reduction and user support team.

\item After each run, we compare the SNR per pixel of the observed spectra with the 
the expected SNR from the ESO exposure time calculator. The SNR per pixel of the
observed spectrum is calculated with the DER-SNR algorithm \citep{Stoehr:2008},
which was developed to perform empirical and unbiased calculations
of SNR on large datasets. When we observe significant discrepancies 
between the expected and the observed SNRs, we investigate if they are
due to the data reduction or to the target selection process. Specifically,
lower than expected SNRs may be due to artificial noise produced by the pipeline, 
low sky transparency, poor accuracy of the stellar astrometry and
photometry. In the upper panels of Fig. \ref{fig:SNR_vs_Mag} we compare the observed and 
predicted SNRs for all the stars with known V magnitude observed during
the first 18 months of the Gaia-ESO Survey. For most of the stars there is a good agreement between the
predicted and the expected SNR, which demonstrates that both the data reduction
and the fiber allocation procedure have been carried out correctly. 
A small number of stars with lower than predicted SNR is expected, since
not all the observations have been performed in conditions of clear sky.
In the bottom panels of Fig. \ref{fig:SNR_vs_Mag} we show the predicted
and observed SNR as a function of wavelength for one star observed
in good weather conditions (ID 18280330+0639516). The plot shows that our method to calculate the SNR
from the observed spectra is consistent with the ESO exposure time calculator.

\end{enumerate}

Our quality control procedure and the fruitful collaboration with the ESO data reduction group 
allowed us to solve most of the problems of the data reduction process.
However, minor issues that need to be solved still affect $\sim4-5$\% of the spectra.
Specifically, the merged spectra of bright stars are affected by ripples 
in the wavelength ranges in common between two orders and, in a small minority of cases, the
sky spectrum below 5000~\AA~is overestimated, so the final spectrum is oversubtracted in
this wavelength range. To avoid that these minor issues may affect the analysis,
the output files include the spectra from each single order and the subtracted sky spectrum.

\section{Data products}

Our products are periodically released to the WGs\footnote{The spectra are released to the other WGs 
of the Gaia-ESO consortium with an operational database developed 
by the Cambridge Astronomical Survey Unit (CASU) based at the Institute of Astronomy 
at the University of Cambridge (see http://casu.ast.cam.ac.uk/gaiaeso/).}, which perform the spectral analysis  
and on different timescales to ESO, which releases them to the general astronomical community by its public website 
(http://www.eso.org/sci/observing/phase3/data\_releases.html).
All our products are organized in multi-extension FITS files. We provide 
three different categories of output files: a) files including all the spectra from a single exposure;
b) files including all the co-added spectra from multiple exposures of the same OB and/or
different OBs with the same stars, same setup, and observed during the same night; c) files including   
only the spectrum of one star, resulting from the stacking of all the spectra of that star
observed with the same setup during the whole survey.    
All the output files include:

\begin{itemize}

\item A spectrum obtained by merging all the spectral orders together, 
before the normalisation, with its variance.

\item A normalized version of the above spectrum, with its variance.

\item All the single spectral order spectra before merging, with their variances.  

\item The spectrum of the sky used for the subtraction.

\item The function used for the continuum normalisation.

\item The CCF used for deriving the RV and $v \sin i$ as discussed
in Sect. \ref{sect:radial_velocities}.

\item A table including information about the stars (e.g. coordinates, name, magnitude),
information about the spectra (e.g. SNR\footnote{In the fits table we report a median of the SNR for the whole wavelength range. 
This value is different from what reported in Fig. \ref{fig:SNR_vs_Mag}, where the SNR is calculated in a small wavelength range to 
allow us a better comparison with the predicted values.}, root mean square of the residuals 
of the wavelength solution), the RV with error, the template used to derive the RV, properties of the CCF (e.g. width and height of the peak), 
$v \sin i$ with error, RV quality flag and binarity flags.

\item The final files with stacked spectra also include the original spectra before co-adding and specific
information on each of them. 

\item Information on the observations (e.g. data, name of the observing block, seeing, airmass) are reported in the header of the files.

\end{itemize}

\section{Summary}

This work is part of a series of papers aimed at describing 
methods, software and procedures used for the Gaia-ESO Survey. Specifically, we 
describe the data reduction and the determination of RVs for the
FLAMES-UVES spectra. 
We can summarize the content of this work as follows:

\begin{enumerate}

\item The basic steps of the data reduction process (bias subtraction, flat-fielding,
wavelength calibration, spectra extraction) are carried out with
a workflow specifically developed to run the ESO public pipeline in the most efficient 
way for the reduction of the spectra of the Gaia-ESO Survey.

\item We perform preliminary steps (i.e. barycentric correction, 
sky-subtraction, spectral co-adding and normalisation) of 
the data analysis with a pipeline based on pyraf and IDL.

\item We derive RVs and $v \sin i$ by cross-correlating 
all the spectra with a sample of synthetic templates. The typical error on 
RVs is $\sigma\sim0.4~\rm km~s^{-1}$ and the major source of error is the variation
of the zero point of the wavelength calibration. A comparison with the RVs measured using 
GIRAFFE spectra indicate the presence of a systematic offset of $\sim0.9~ \rm km~s^{-1}$ 
between the two instruments.
We are investigating how to improve the precision and the accuracy of the RVs (e.g. by using sky lines) 
and we are carrying out an overall assessment of the zero point shifts of all the instruments and setups 
to put all the RVs of the Gaia-ESO Survey on the same zero point.

\item We perform a detailed quality control of our final products, which is based on
the analysis of the output parameters from the ESO pipeline, a visual inspection of the spectra
and an analysis of the SNR.

\item Our output is organized in multi-extension FITS files, which include both spectra
at various stages of the data reduction process (e.g. normalized, not normalized, co-added, not co-added),
and various information on the spectra, which are collected in tables (e.g. coordinates, RVs, magnitudes).   

\end{enumerate}
  
\begin{acknowledgements}
We acknowledge the support from INAF and Ministero dell'Istruzione, dell' Universita' e 
della Ricerca (MIUR) in the form of the grant "Premiale VLT 2012".
The results presented here benefit from discussions held during the 
Gaia-ESO workshops and conferences supported by the ESF (European Science Foundation) through 
the GREAT Research Network Programme.
This work was partly supported by the European Union FP7 
programme through ERC grant number 320360 and the Leverhulme Trust through grant RPG-2012-541.
We acknowledge financial support from "Programme National de Cosmologie and Galaxies" 
(PNCG) of CNRS/INSU, France.
\end{acknowledgements}

\bibliographystyle{aa}
\bibliography{/Users/sacco/LETTERATURA/bibtex_all}

\end{document}